\documentstyle[12pt]{article}        
\begin{document}
\setlength{\textwidth}{15truecm}
\setlength{\textheight}{23cm}
\baselineskip=18pt
\bibliographystyle{unsrt}
\begin{center}
{\Large{\bf Dynamics of Forster Energy Migration Across Polymer Chains 
in Solution}} \\
\vspace{1cm}
{\large{\bf G. Srinivas, A. Yethiraj \footnote [1] {Permanent address:
Department of Chemistry, University of Wisconsin, Madison, 
WI 53706, USA. E-mail: yethiraj@chem.wisc.edu}$^,$\footnote [2] 
{Also at the 
Jawaharlal Nehru Center for Advanced Scientific Research, jakkur, Bangalore,
India;}  and B. Bagchi}} 
\footnote[3] {Corresponding author. 
E-mail: bbagchi@sscu.iisc.ernet.in}$^{,\large \bf \star}$\\
\vspace{0.5cm}
\vspace{0.1cm}
Solid State and Structural Chemistry Unit,
Indian Institute of Science,
Bangalore, India 560 012.\\

\end{center}
\vspace{0.3cm}
\baselineskip=18pt
\begin{center}
{\bf Abstract}
\end{center}
\vspace{-0.4cm}
{\small 
\baselineskip=18pt

 Long distance excitation energy transfer between a donor and an 
 acceptor embedded in
a polymer chain is usually assumed to occur
via the Forster mechanism which predicts a $R^{-6}$ distance dependence  
of the transfer rate, where $R$ is the distance between the donor
and the acceptor. In solution $R$ fluctuates with time.
In this work, a Brownian dynamics 
simulation of a polymer chain with  Forster energy transfer between the
two ends is carried out and the time dependence
of the survival probability $S_{P}(t)$  is obtained. The latter can be measured
by fluorescence resonance energy transfer (FRET) technique which is now widely
used to study conformations of biopolymers 
via single molecule spectroscopy. It is found that 
the survival probability is exponential-like when the  Forster radius ($R_F$)
  is comparable to the root mean square radius ($L$) of the polymer chain.
 The decay is strongly non-exponential both for small and large $R_F$, and
 also for large $k_F$. Large deviations
from the Wilemski-Fixman theory is obtained when $R_F$ is 
significantly different from $L$.


\baselineskip=22pt

\section{Introduction}

     Fluorescence resonance energy transfer (FRET) is a powerful
  technique to study many aspects of structure and dynamics of polymers
  and biopolymers.$^{1-8}$ In this technique, the polymer is doped with a donor
  and an acceptor at suitable positions along the chain. The donor 
  is initially excited optically by a laser light which subsequently 
  transfers its energy to the acceptor which is located at
  a distance R from the donor. In many applications, the distance between
   the donor and the 
  acceptor is fixed, as in the case of rigid biopolymers.$^{4,5,7,8}$
   However, in many
  cases of interest, this distance is a fluctuating quantity.$^{1-3,6}$
   For example,  the distance between the two ends of a polymer in solution
   executes a Brownian motion. If the polymer
  is assumed to be an ideal Gaussian chain, then the 
  mean square distance between
  the two ends is $Nb^{2}$, where $N$ is the number of the monomer units and 
  $b$
  is monomer (or the Kuhn) length. Thus, as the number of monomers in the
  chain increases, the average distance between the two ends also increases.
  The probability distribution of $R$, $P(R)$, however, remains peaked at
  $R=0$ for ideal chain, although its height decreases as $1/\sqrt(N)$. All
  these properties play important parts in determining the observed dynamics
  of excitation energy transfer.

    The usually assumed mechanism 
    for excitation energy transfer between a donor and 
    an acceptor is the Forster mechanism$^{6}$ which gives the following
    expression
    for the  singlet-singlet resonance energy transfer rate, $k(R)$,
\begin{equation}    
	k(R) =k_{F} \frac{1}{1 + (R/R_{F})^{6}}
\end{equation}
where $R_{F}$ is the Forster
radius and $k_{F}$ is the rate of excitation transfer when the separation
between the donor and the acceptor is very small, that is $R/R_F\rightarrow0$.  
The Forster radius is usually obtained from the overlap of the donor 
fluorescence with the acceptor absorption and several other available
parameters$^5$.
  
   The dynamics of Forster energy migration has been traditionally investigated
  by performing time domain measurements of the decay of
  the fluorescence (due to excitation transfer) from the donor.$^{1,4-6}$ 
  More recently, this technique
  has been used in single molecule spectroscopy of biopolymers$^{7,8}$.
  In the latter, distance dependence of FRET provides relevant information
  about the conformation and dynamics of single biopolymers. Recently,
  FRET from single protein molecules has also been used to study 
  protein folding$^9$.
   At any given time after the initial
  excitation, the fluorescence intensity is a measure of the "unreacted"
  donor concentration. As both $k_F$ and $R_F$ are determined
  by the donor-
  acceptor pair, the rate of decay of the fluorescence intensity 
   provides a direct probe of the conformational dynamics of the polymer.
   
  When the polymer is in solution, each monomer (or polymer bead) executes Brownian
  motion. Because of the connectivity of the polymer chain, this Brownian
  motion of the monomers are strongly correlated. This many body nature
  of polymer dynamics  can be described by a joint,
  time dependent  probability distribution $P({\bf r}^N,t)$ 
  where ${\bf r}^N$ denotes the positions of all the N polymer beads.
  The time dependence of the   probability distribution $P({\bf r}^N,t)$
  can be described by the following
  reaction-diffusion equation$^{10,11}$
  
\begin{equation}
{\partial \over \partial t } P({{\bf r}^N},t) = {\cal L}_B({{\bf r}^N},t)
P({\bf r}^N,t)  
  - k(R) P({{\bf r}^N},t)
\end{equation} 
\noindent where ${\cal L}_B$ is the full $3N$ dimensional diffusion 
operator given by,
\begin{equation}
{\cal L}_B({{\bf r}},t)= D \sum^{N}_{j=1} 
{\partial \over \partial r_j} P_{eq}({{\bf r}},t)
{\partial\over \partial r_j} {P({{\bf r}},t) \over P_{eq}({{\bf r}},t)}
\end{equation} 
$R$ is the scalar distance between the two ends of the polymer chain and $D$ is
the center of mass diffusion coefficient.

  The solution of Eq.2 with the sink term given by the Forster expression
is highly non-trivial. In two seminal papers, Wilemski and Fixman presented
a nearly analytic solution of the problem for any arbitrary sink.$^{10,11}$
 The WF theory
has been tested ($\it only$ for the average rate) by computer simulations
when the sink is a Heaviside
function.$^{11-13}$ We are not aware of any such simulation 
study with the Forster rate. Such a study is clearly important.

   The objective of this paper is mainly two fold. First, to present
   results of detailed Brownian dynamics (BD) simulations of Eq.2, with k(R)
   given by the Forster rate (Eq.1). Detailed investigation into the time
   dependence of the survival probability at a time t after the initial
   excitation ($S_{P}(t)$) is presented. We believe that this is the first
   such calculation of $S_{P}(t)$ for this kind of a problem. We find that
   the time dependence of $S_{P}(t)$ can be non-exponential for a large
   range of the relevant parameter space (N,$k_F,R_F$).
   Second, we present a detailed comparison
   of the simulated rate with the WF theory. This comparison has been carried
   out at the level of time dependent survival probability, again, we believe,
   for the first time.
   
   The organization of the rest of the paper is as follows. In next section, 
   we present the details of the BD simulation. In section III, we discuss the
   WF theory. In section IV we present the simulation results and the comparison
   with the WF theory. Section V concluded with a discussion.
  
\section {Simulation Details}

	Brownian Dynamics (BD) simulations of polymer motion have been 
	carried out with an 
idealized Rouse chain, in which the set of beads are connected by the 
harmonic potential,
\begin{equation}
\beta U = {3\over 2 b^2} \sum_{j=1}^{N-1} (r_j - r_{j+1})^2
\end{equation} 
\noindent where the position vector of a bead $j$ is denoted by $r_j$, 
and $N$ is the number of beads constituting the 
polymer chain. The mean 
squared bond length is $b^2$. Equilibrium end-to-end distance of the 
polymer chain is,         
\begin{equation}
\langle(r_N - r_0)^2 \rangle = L^2 = Nb^2
\end{equation} 
In Rouse model, there exists no excluded volume forces and the 
hydrodynamic interactions between the monomer beads are ignored$^{13}$.
More details on this model can be found elsewhere$^{13}$.
In our present study the polymer chain is additionally characterized by the
presence of two reactive end groups. This essentially implies that within 
the time interval $\Delta t$, the
two end groups react with a  probability $k(R) \Delta t$. 

  The initial configuration for each trajectory has been selected from 
a Monte Carlo generated random distribution. The following equation of
motion has been used in the simulations,
\begin{equation}
r_j(t+\Delta t) = r_j(t) + F_j(t) \Delta t + \Delta X^G(t)
\end{equation} 
\noindent where the positions of the j-th bead at time $t$ and $t+\Delta t$ 
are denoted by 
$r_j(t)$ and $r_j(t+\Delta t)$, respectively. $F_j(t)$ is the total
force acting
on bead $j$ and $\Delta X^G(t)$ is a random Brownian displacement, taken
from a Gaussian distribution with zero mean and the variance 
$\langle (X^G)^2 \rangle = 2 \Delta t$. In writing Eq.1, we have set
both $D_{0}$ and $K_{B}T$ equal to unity; the latter is Boltzmann constant
times the temperature. Here we adopted a time step,
$\Delta t = 0.001$. However, smaller time steps as low as 0.0001 has been 
employed in the limit of both large $k_F$ and large $R_F$ values, to account for
the faster dynamics. During the simulation both the 
 mass $m$ and the root mean-square bond length
$b^2$ of the bead have been set to unity for computational convenience.

   The  trajectory
generated by using the above procedure needs to be
 terminated when the two end groups react. 
This has been done in simulations in the following way. Each time
the trajectory is updated, the existing end to end distance $R$ is used
in Eq.1 to calculate the distance dependent rate constant, $k(R)$. We then
call a random number generator to get a value between zero and unity. If
this value is less than $k(R) \Delta t$, then the trajectory is terminated.
Otherwise, the trajectory is continued. One forms a histogram over many
such trajectories. This procedure generates an irreversible FRET.

For each polymer chain constructed randomly, 
we have equilibrated it for 10,000 time steps before 
switching on the reaction. Subsequently, 50,000 to 
1 Lakh trajectories with different initial configurations were generated
 and
the survival probability $S_p(t)$ was obtained by averaging over all the 
trajectories.  This procedure was systematically applied for
the polymer chains containing the beads, $ N= 20, 50 $ and $100$.

  Before proceeding with the simulations of the Forster transfer, we reproduced
 the results of Pastor,Zwanzig and Szabo (PZS)$^{12}$ on the mean first passage time
 with Heaviside sink function of infinite strength. Our simulation results
 agreed with  those of PZS within the uncertainity given by PZS.

\section {Wilemski-Fixman Theory $^{10,11}$}
  
  Several decades ago Wilemski and Fixman (WF) developed a non-trivial
theory for the diffusion 
limited intrachain reaction of a flexible polymer.$^{10,11}$ To account for the chemical 
reaction they have added a sink term ${\cal S}$ to the manybody 
diffusion equation, just as in Eq.2. The WF equation of motion is well-known
and we present it below for the sake of completion
\begin{equation}
{\partial \over \partial t} P({\bf{r}}^N,t) + {\cal L}_B P({{\bf r}^N},t) 
= -k_0 {\cal S}(R) P({{\bf r}^N},t).
\end{equation} 
\noindent In the notation of the present work
\begin{equation}
k_0= k_F; S(R) =  \frac{1}{1 + (R/R_{F})^{6}}. 
\end{equation} 
\noindent The operator 
${\cal L}_B({\bf r}^N,t)$ is given by Eq.3. 
As already mentioned, the treatment of WF is general and can be applied
to a reaction with arbitrary sink.

 Let us define a survival  probability $S_p(t)$ as 
the probability that the
chain has not reacted after time $t$. $S_p(t)$  is then given by,
\begin{equation}
 S_p(t) = \int P({\bf r}^N,t) d{\bf r}_1 d{\bf r}_2...d{\bf r}_N
\end{equation} 
 In order to obtain the survival probability WF made a closer approximation, 
according to which the Laplace transform 
of $S_P(t)$ can be written as, 
\begin{equation}
{\hat S}_p(s)= {1 \over s} - {k \upsilon_{eq} \over s^2 
(1+ {k \hat D}(s)/\upsilon_{eq})}
\end{equation}
\noindent where ${\hat D}(s)$ is defined as, 
\begin{equation}
{\hat D}(s) = \int_0^\infty e^{-st} D(t) dt
\end{equation} 
\noindent which is the Laplace  transform of sink-sink 
time correlation function $D(t)$ defined as,
\begin{equation}
D(t) = \int d^3{\bf R}_1\int d^3{\bf R}_2  {\cal S}({\bf R}_1)  
{\cal S}({\bf R}_2) G({\bf R}_1,{\bf R}_2,t)
\end{equation} 
\noindent The Green function appearing in the above equation is
given by,   
\begin{eqnarray}
G({\bf R}_1,{\bf R}_2,t) = \biggl({3\over 2 \pi L^2} \biggr)^{3/2}
\biggl( {1\over (1-\rho^2)^{3/2}} \biggl) 
exp\biggl(-{R_{1}^2 - 2 \rho(t) {\bf R}_1.{\bf R}_2 + R{_2}^2 
\over 2 L^2 (1- \rho^2)} \biggr)
\end{eqnarray}  
\noindent where $\rho (t)$ is the normalized time correlation function of 
end-to-end vector $\langle {\bf R}(0).{\bf R}(t) \rangle/\langle {R}^2
\rangle $ which can be obtained analytically and is given
 by the following equation,
\begin{equation}
\rho(t) = {8 \over \pi^2} \sum_{l; odd} {4 \over l^2} exp(-\lambda_l t)
\end{equation} 
If we neglect excluded volume and hydrodynamic interactions,  
$\lambda_l$ is given by the following expression$^{10,11}$.
\begin{equation}
\lambda_l = 3D_{0}(l\pi/Nb)^2,
\end{equation} 
 Finally  ${\upsilon_{eq}}$ is given as,
\begin{equation}
\lim\limits_{t\rightarrow \infty} D(t) = (\upsilon_{eq})^2
\end{equation} 
Note that $\upsilon_{eq}$ is the rate when the distribution of the polymer
ends is at equilibrium. Thus, $\upsilon_{eq}$ gives the initial rate of decay
of $S_{P}(t)$ which will show up as the transient behavior. 
In most cases, the rate of decay should become progressively slower, as
the population from the sink region decreases as the reaction proceeds.

 Once the choice of sink function specified, it is straight forward to 
calculate
the survival probability by utilizing the above set of equations. WF choice was
the heaviside sink function. Later, 
Doi showed that WF method is easy to apply if the heaviside sink function is 
replaced with
a Gaussian sink function$^{14}$. Bettizzeti and Perico studied the 
dependence of the rate on
the choice of sink function with in the frame work WF theory and supported the
WF closure approximation$^{15}$. 
Surprisingly, no analysis of the time dependence of the survival probability,
$S_{P}(t)$ has ever been reported.
In this study we follow the orignal scheme
proposed by WF to obtain $S_p(t)$ analytically. In doing so we use the
Stefest algorithm to obtain $S_p(t)$ through the Laplace inversion of Eq. 10.

\section{Results and Discussion}

   Before discussing the results let us describe the scaling that has been 
used to compare the results obtained by simulation with the theory.
In the reduced unit notation adopted in simulation, the rate constant 
has been scaled as $\tilde k_F=k_Fb^2/D_{0}$ and the real time has been scaled by 
$b^2/D_0$. However, in the original WF theory, 
time is scaled by $6D/L^2$ where $D$ is the center of mass diffusion.
In the free draining limit, so $D=D_{0}/N$ and  $L^{2}=Nb^2$. Thus, 
the numbers obtained from WF theory is to be converted to the 
simulation scaling for a comparison of results. The Forster radius
is scaled by the bead diameter, $b$. Another important parameter in this problem
is the root mean square radius of the polymer as this determines the end to end
distribution. Although we have carried out simulations for N=20, 50 and 100,
in this report we  shall concentrate mostly on N=50.
  
	Figures 1 and 2 depict the time dependence of the survival probability
$S_p(t)$ for two different values of $R_F$, $R_F=1$ and $R_F=5$,
respectively, for a fixed $N=50$. In  figure 1, $k_F$ has been varied
from 50 to 0.1 , that is, over two orders of magnitude. The decay remains
non-exponential over the whole range. In figure 2, $k_F$ has been varied from
10 to 0.1. Here the decay is exponential-like. These two figures demonstrate
the strong dependence of the decay profile of $S_{P}(t)$ on $R_F$. Note that 
the earlier experiments which fitted the quantum yield to the Forster expression
obtained values which are rather large, comparable to the ones shown in
figure 2. This could have been due to the use of an equilibrium end-to-end
distribution
in the fitting, instead of a time dependent probability distribution. 
In model calculations,  one usually assumes a small value of $R_F$ (
often in the form of a Heaviside sink function). This
strong dependence of decay profile on $R_F$ could be potentially useful
in unravelling mechanism and dynamics of energy transfer.

  It is not difficult to understand the above results qualitatively. For
  an ideal Gaussian chain, the maximum in the probability distribution that the
  two ends are separated by a distance R is located at $\sqrt(2N/3)b$. For
  N=50, this value is 5.773b. Therefore, when $R_F$ is equal to 5, the decay
  is facilitated by the presence of a large fraction of the distribution at
  a distance of separation where the transfer rate is large. This can explain
  the exponential-like decay for $R_F=5$ (figure 2). However, the situation
  is completely different for $R_F=1$. Here the probability of finding a
   polymer with end to end distance so small is negligible {\em and}
   the transfer rate where
   the bulk of the population is located is very small because of the strong $R$ 
   dependence of the Forster transfer rate. 
   Therefore, the decay of the survival
   probability starts slowly (Fig.1) and is determined by the interplay between
   the diffusion and the rate. This explains the shape of figure 1.
  
   The above discussion also suggests that the shape of the survival probability
   can depend rather strongly on the length of the polymer chain. This is
   because the Forster distance for a given donor-acceptor pair is likely to be
   independent of the length of the polymer chain. But the distribution and also the
   diffusion rate will be determined by N. However, this dependence is not
   trivial and will be discussed elsewhere.

    In figures 3 and 4, we have compared predictions of the WF theory with the
    simulations. In figure 3, $S_{P}(t)$ is plotted for two very different
    values of $k_F$ ($K_F =$ 1 and 10) at $R_F=5$ for $N=50$. 
    It is seen that while the agreement 
     is satisfactory at short times for both the cases, 
     the same is not true at long times, particularly for the smaller
     $k_F$. In the latter case the simulation also finds a larger 
     non-exponentiality, as discussed later.
     In figure 4, $S_{P}(t)$ is plotted for 
    for $R_F=1$ , $k_{F}=$1.0 (and $N=50$). It is seen that the
    WF theory breaks down in this limit. This is one of the main results
    of the present study. The agreement improves
    for smaller $K_F$ but becomes worse in the opposite limit.
    
    We have also simulated both larger and smaller chains.
    Since the size of the
    polymer scales with N, it is not possible to compare  results
    for different sizes. In figure 5, we show the comparison between the
    simulation results and the WF theory for $N=100$ at $R_{F}=8$ and
    $k_{F}=0.1$. This is the most favorable parameter space for
    the WF theory. Although the simulated $S_{P}(t)$ decays somewhat faster,
    the WF theory prediction is not totally off. We have not yet 
    searched for any scaling laws (in N, $R_F$ and L) dependence
     -- work is under progress in this direction.

     In figures 6 and 7, we present logarithmic plots of $S_{P}(t)$
     to show the extent of non-exponentiality, for different values of $R_F$ 
     and $k_F$. It can be seen that while the decay is nearly exponential for
     small $K_F$ (which is expected), 
     it is strongly non-exponential when the value of $k_F$
      becomes comparable to or larger than the bead diffusion rate, $D_{0}/b^2$.

    The inability of the Wilemski-Fixman theory to explain
    the time dependence of the survival probability is surprising. We note
    that earlier theoretical studies have considered only the mean first 
    passage time. In figure 8 we have compared the simulated end-to-end
    vector time correlation function ($\rho(t)$) with the slightly approximate
    expression used by WF. The  agreement is good, as expected. This
    agreement improves further for larger N. Thus, the failure of WF 
    (as shown in Fig.4) must be due to the closure approximation.

\section{Conclusions}

      Use of FRET in single molecule spectroscopy of polymers and biopolymers
  requires accurate knowledge of the mechanism of energy transfer, more
  importantly, the distance dependence of the transfer rate. The fluorescence
  quantum yield can provide only  an average estimate of the distance between
  the donor and the acceptor if the mechanism is well-understood. This could
  be sufficient for rigid systems. For many systems of interest, for example
  for understanding the dynamics of protein folding or in the collapse of
  polymers, one requires the time dependence of the excitation migration. This
  will be measured in terms of the time dependent survival probability.
   
  In this work we have presented results of detailed Brownian dynamics
  simulations of Forster energy transfer between the two ends of an
  ideal Gaussian chain. As noted by previous workers$^{12}$, this apparently
  simple problem is actually highly non-trivial because this is a manybody
  problem. We have calculated survival probability for a large number of values
  of the transfer rate $k_F$ and the Forster radius, $R_F$. It is found that 
  while the survival probability is exponential-like for small values of $k_F$
  (compared to the monomer diffusion rate, $D_{0}/b^2$) {\it and} intermediate
   $R_F$, it is strongly non-exponential for small (compared to L) $R_F$.
  
   We have compared the results of the simulation with the well-known theory of
   Wilemski and Fixman. It is found that the theory is reliable when the
   Forster radius $R_F$ is comparable to the root mean square radius L
   of the polymer chain and the transfer rate $k_F$ is comparable to or
   smaller than the monomer diffusion rate $D_{0}/b^2$.
   However, the agrement is not at all satisfactory in the limit when $R_F$ 
   is much smaller or
   larger than L. 
   
    What is the reason for the failure of the WF theory when $R_F$ is 
  substantially different from the root mean square radius of the 
  polymer? While it is obvious that
    the WF closure approximation is inadequate in many situations, 
    the exact reason for the failure is not clear. In fact, for the
    nature of the decay curve for large or small $R_F$ can possibly be
    understood even from a one dimensional theory, provided the end to end
    distance correlation function $\rho(t)$ is given. This problem will
    be discussed elsewhere.\cite{gsbb}
    
    Note that the distance dependent rate appears in several other chemical
    processes, like in electron transfer reactions where the rate of transfer
    is known to show an exponential distance dependence. It will be
    interesting to study
    this problem with the method employed here. Another important, long
    standing problem is the study of reactions in realistic polymer
    chains with excluded volume
    and hydrodynamic interactions. Work in this direction is under
    progress.

{\bf \large Acknowledgement}

   This work is supported in parts by the Council of Scientific and Industrial
   Research (CSIR) and the Department of Science and Technology,
   India. The visit of A. Yethiraj was supported by the Jawaharlal Nehru Center
   for Advanced Scientific Research, Bangalore, India. G. Srinivas thanks
   CSIR, New Delhi, India for a research fellowship.

\newpage

\noindent {\bf \large Figure captions:}

\noindent{\bf Figure 1.} The survival probability obtained from Brownian
dynamics (BD) simulations of Eq.2
is plotted against the scaled time  for several values of $k_F$ at $R_F=$ 1. 
The 
curves from top to bottom represent the cases with $k_F=$ 0.1, 1, 10 and 50, 
respectively.

\noindent{\bf Figure 2.} The survival probability $S_p(t)$, obtained from
BD simulations is plotted for $k_F=$ 0.1, 1 and 10 at $R_F=$ 5. Curves 
from top to bottom show  $S_p(t)$ at $k_F=$ 0.1, 1 and 10, respectively.

\noindent {\bf Figure 3.} BD simulation results have been
 compared with WF theory
at a Forster radius, $R_F=$ 5. The upper 
set shows the case with $k_F=$ 1 and the lower set is for $k_F=10$. In both
the cases, symbols shows  the simulation results while the WF theory
 predictions are
represented by the full lines.

\noindent {\bf Figure 4.} WF theory has been compared with the 
simulation results at
a lower value of $R_F$, namely $R_F=1$ and for $k_F$=1. WF theory prediction
has been shown by the full line while the symbols represent the simulation results. 
As seen from 
the figure, WF theory seems to break down in this limit.

\noindent {\bf Figure 5.} The comparison between WF theory and simulation results
has been shown  for a larger polymer chain, $N=100$, for $R_{F} = 8 $ and
$k_F$ $= 0.1$. Symbols and the
full line represents the results of simulation and WF theory, respectively.

\noindent {\bf Figure 6.} The semilog plot of the survival probability 
$S_p(t)$ which has been obtained
from simulations, is plotted against the scaled time at $k_F=$ 0.1 and $R_F=$ 5. 
This 
figure shows that the decay is nearly  exponential for $R_F=$ 5 and small
$k_F$.                                 

\noindent {\bf Figure 7.} The semilog plot of the survival probability,
obtained from simulations is plotted against the scaled time at $k_F=$ 1.0 and 
$R_F=$ 1. Highly non-exponential behavior of $S_p(t)$, 
is very clear in this limit.

\noindent {\bf Figure 8.} The end-to-end vector time correlation function
$\rho (t)$ is plotted against the scaled time for a polymer of mean square 
length $L^2 = 50 b^2$.
Symbols shows the simulated $\rho (t)$ while the $\rho (t)$ obtained from WF 
expression is shown by full line.

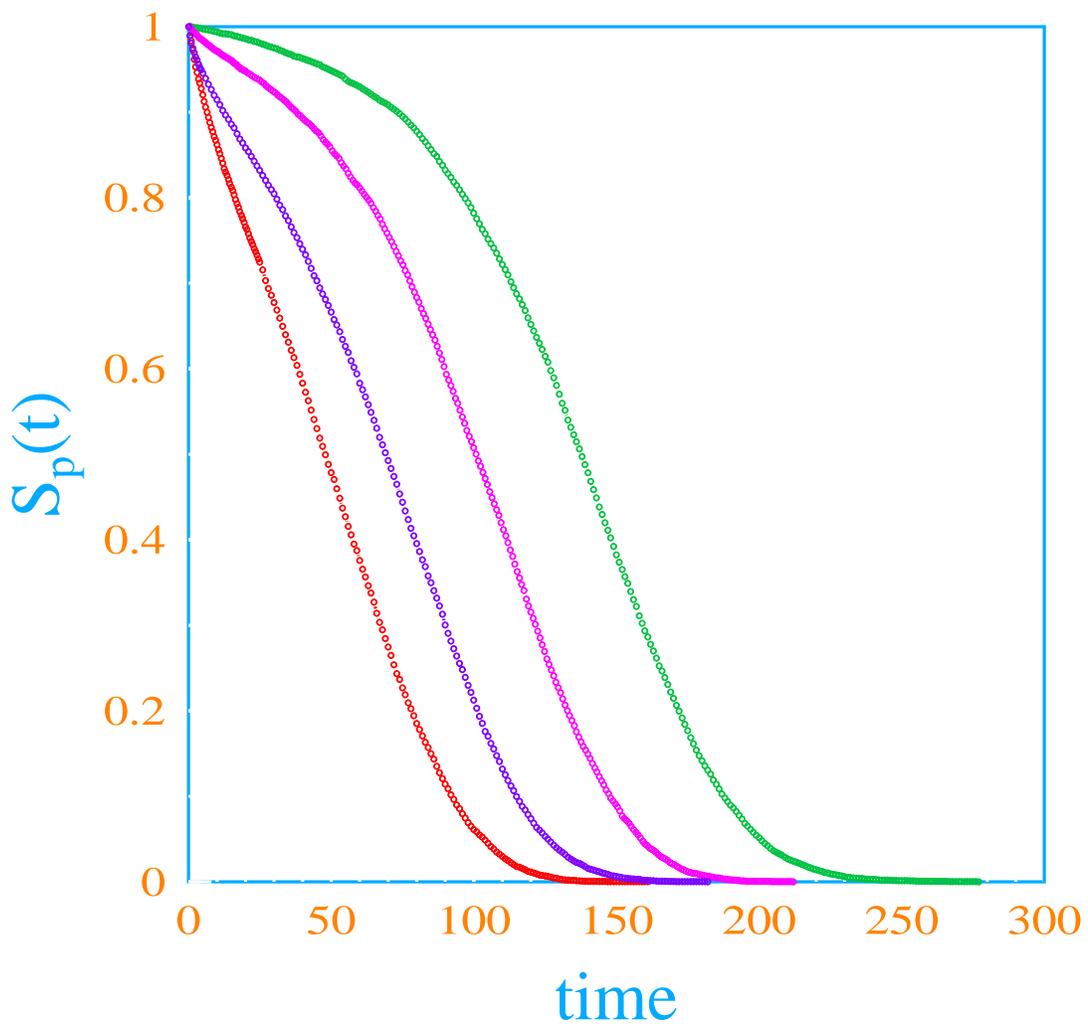
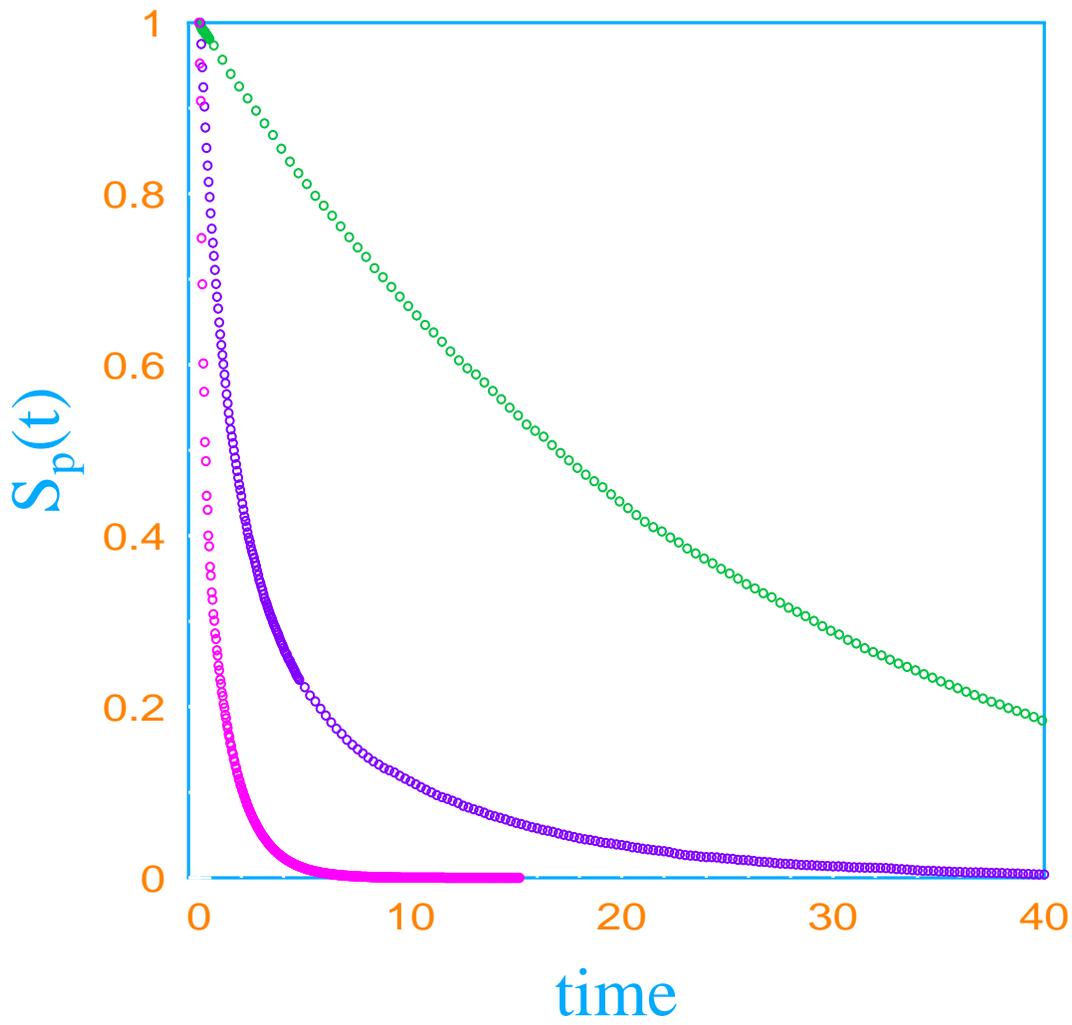
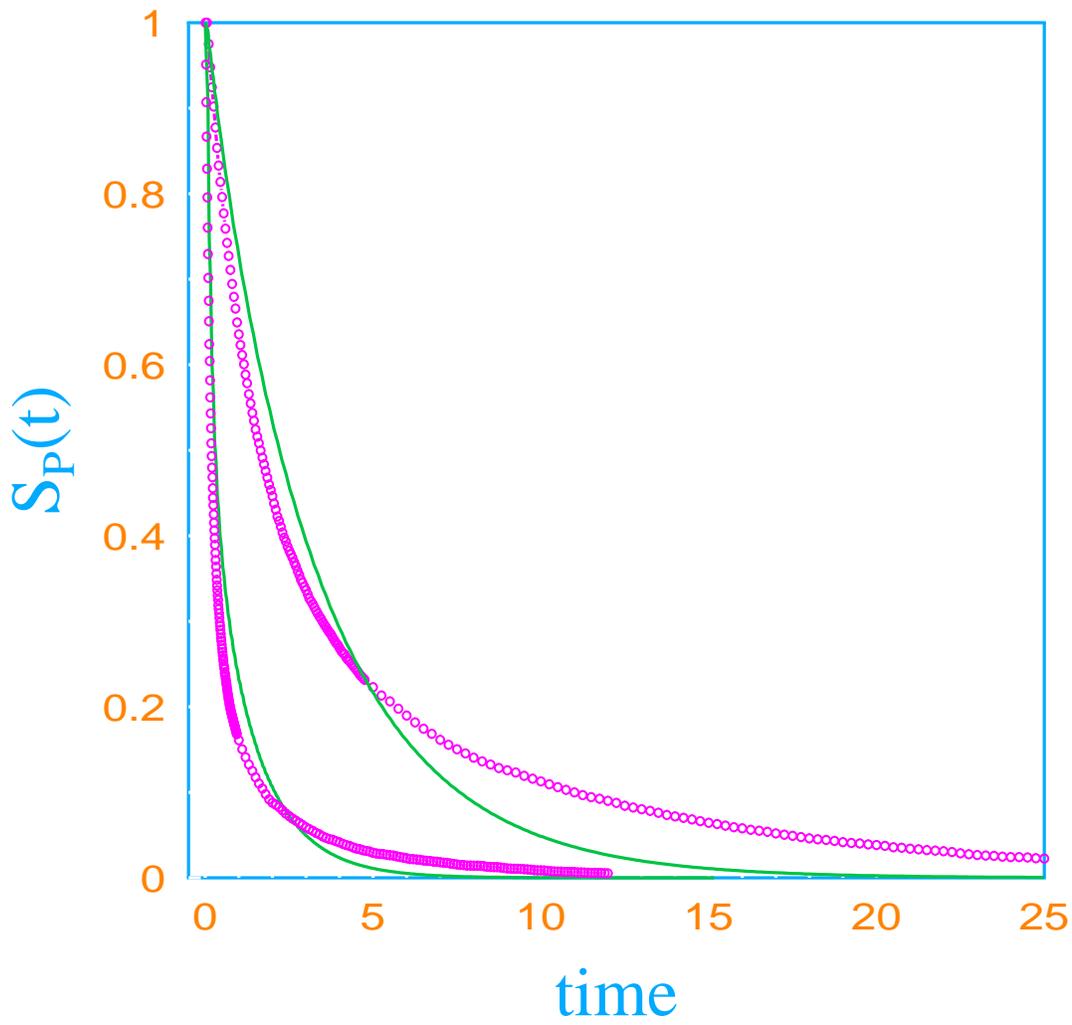
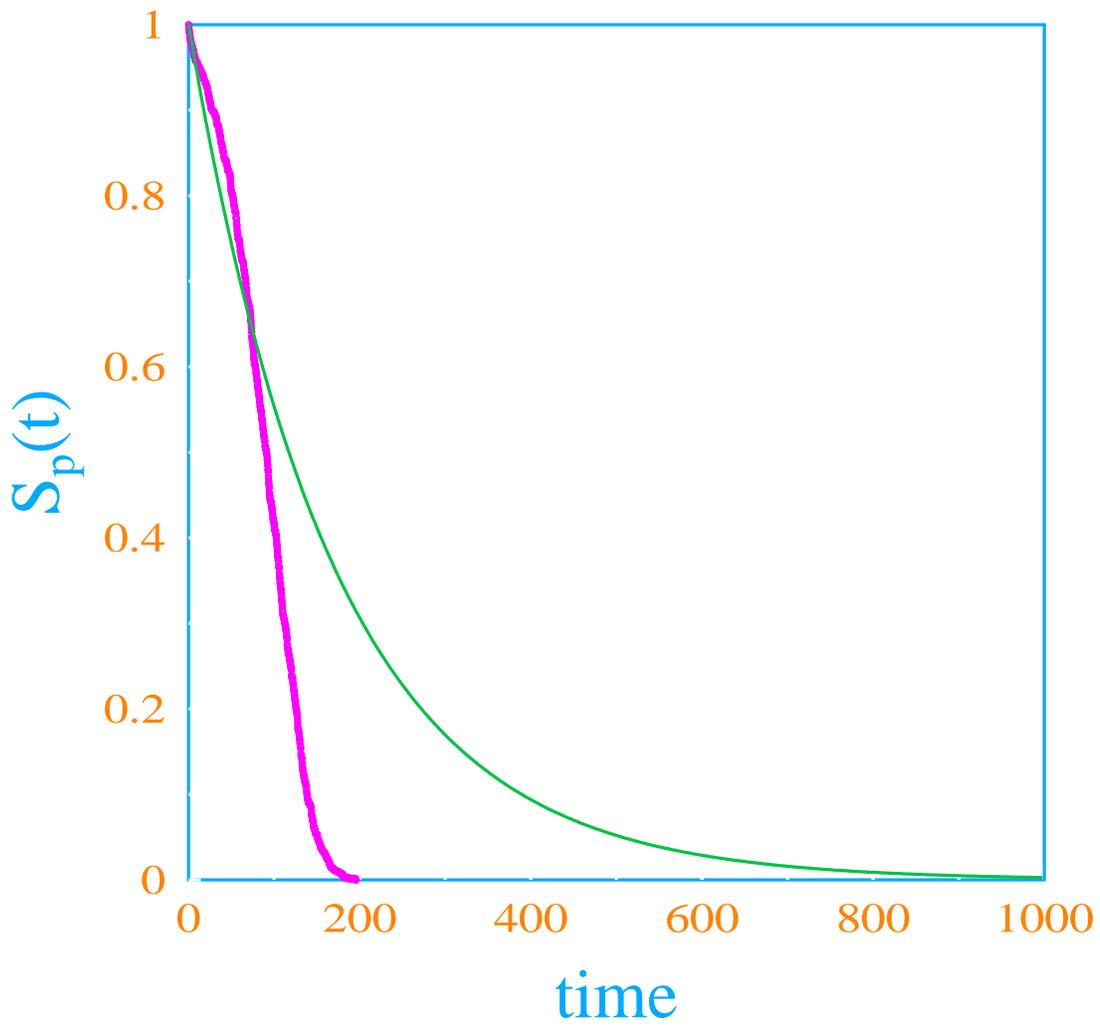
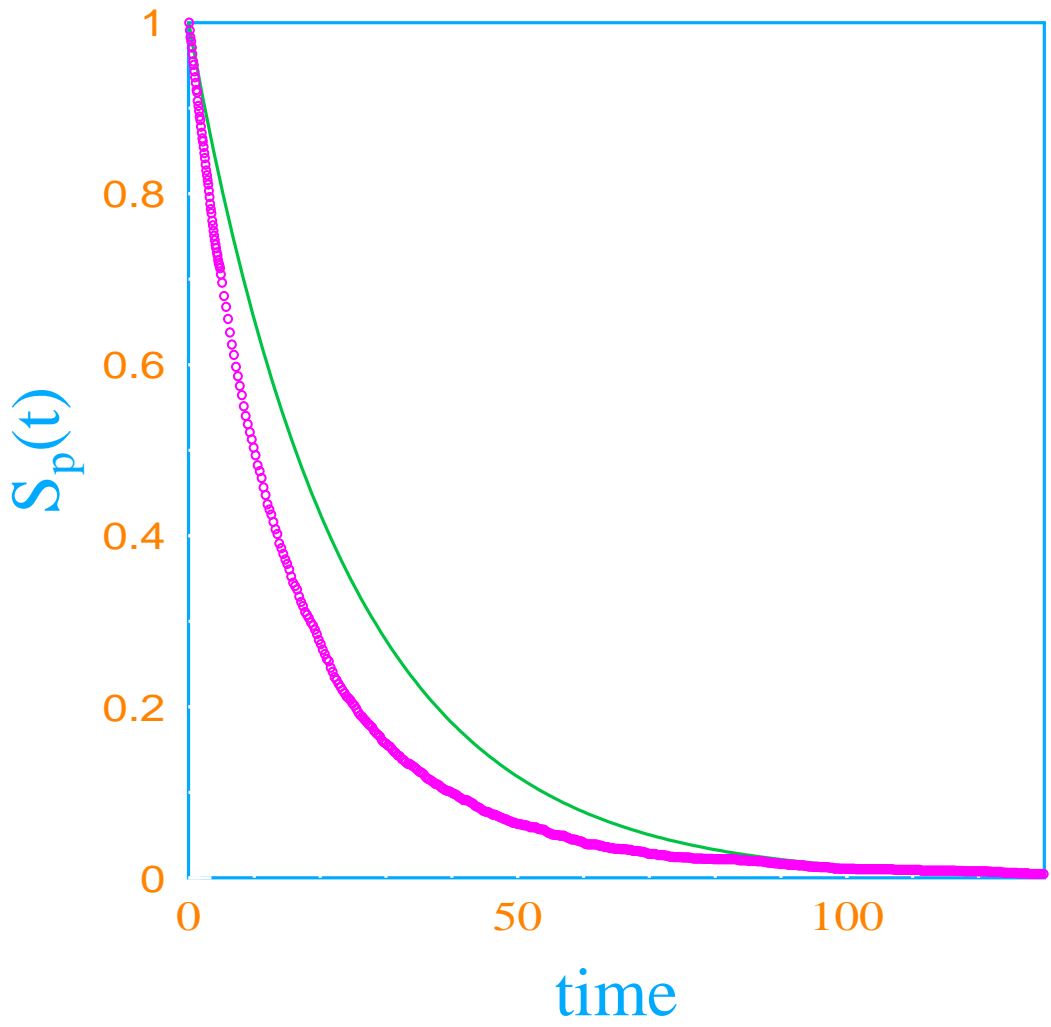
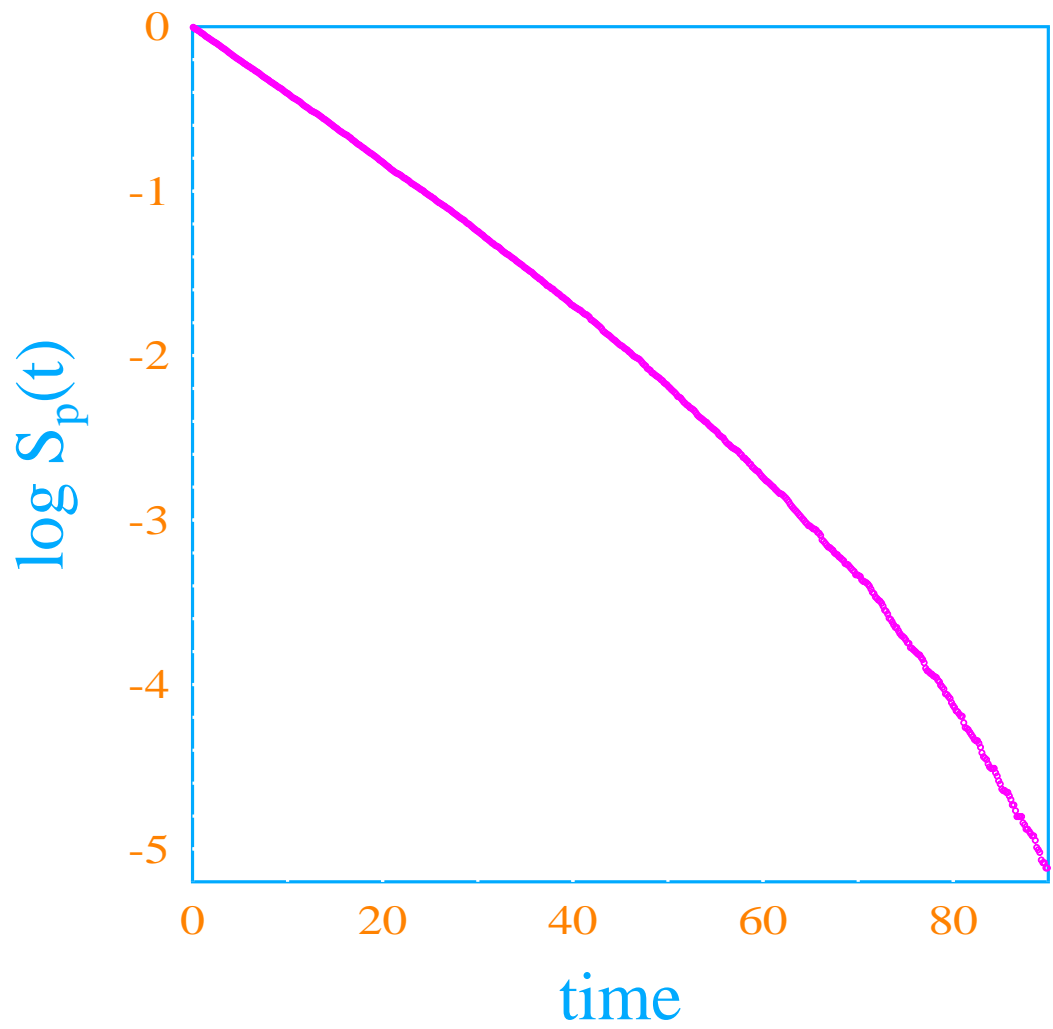
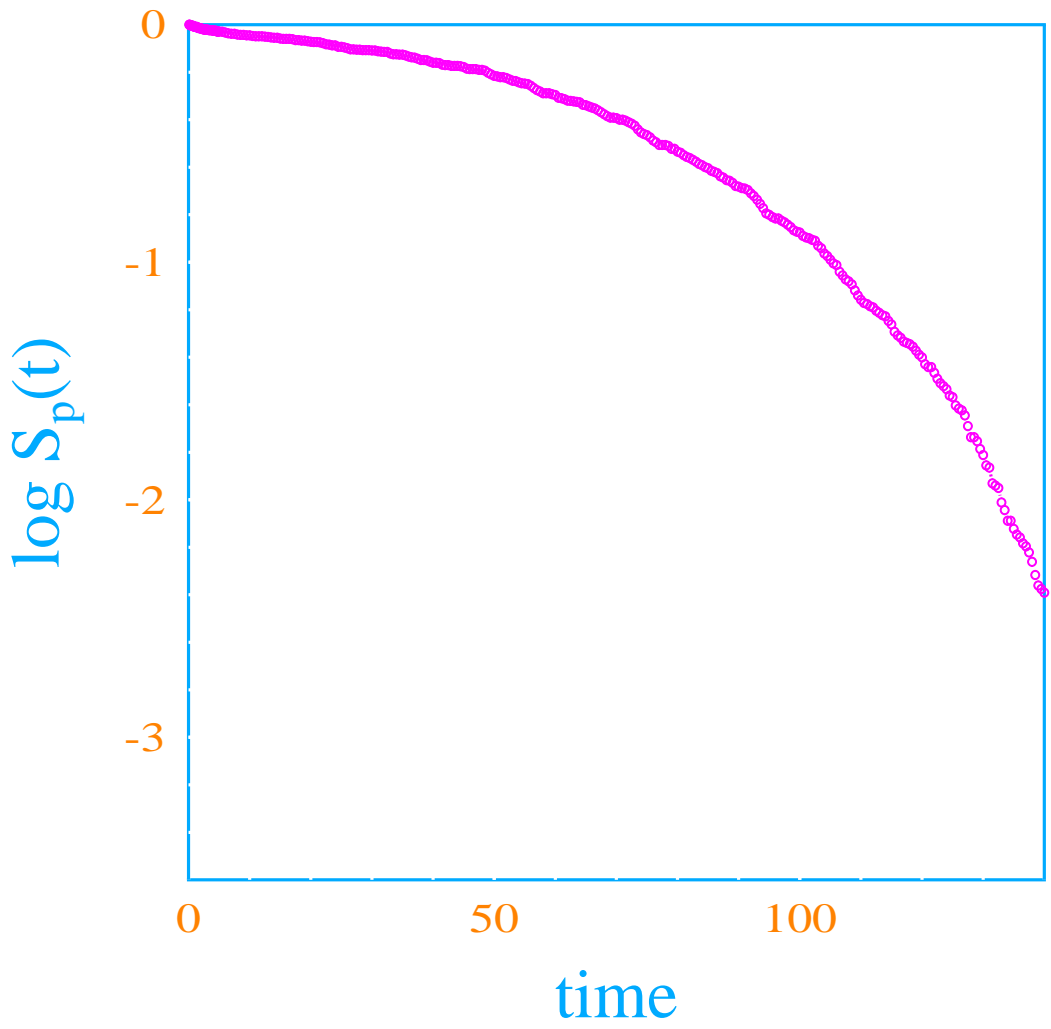
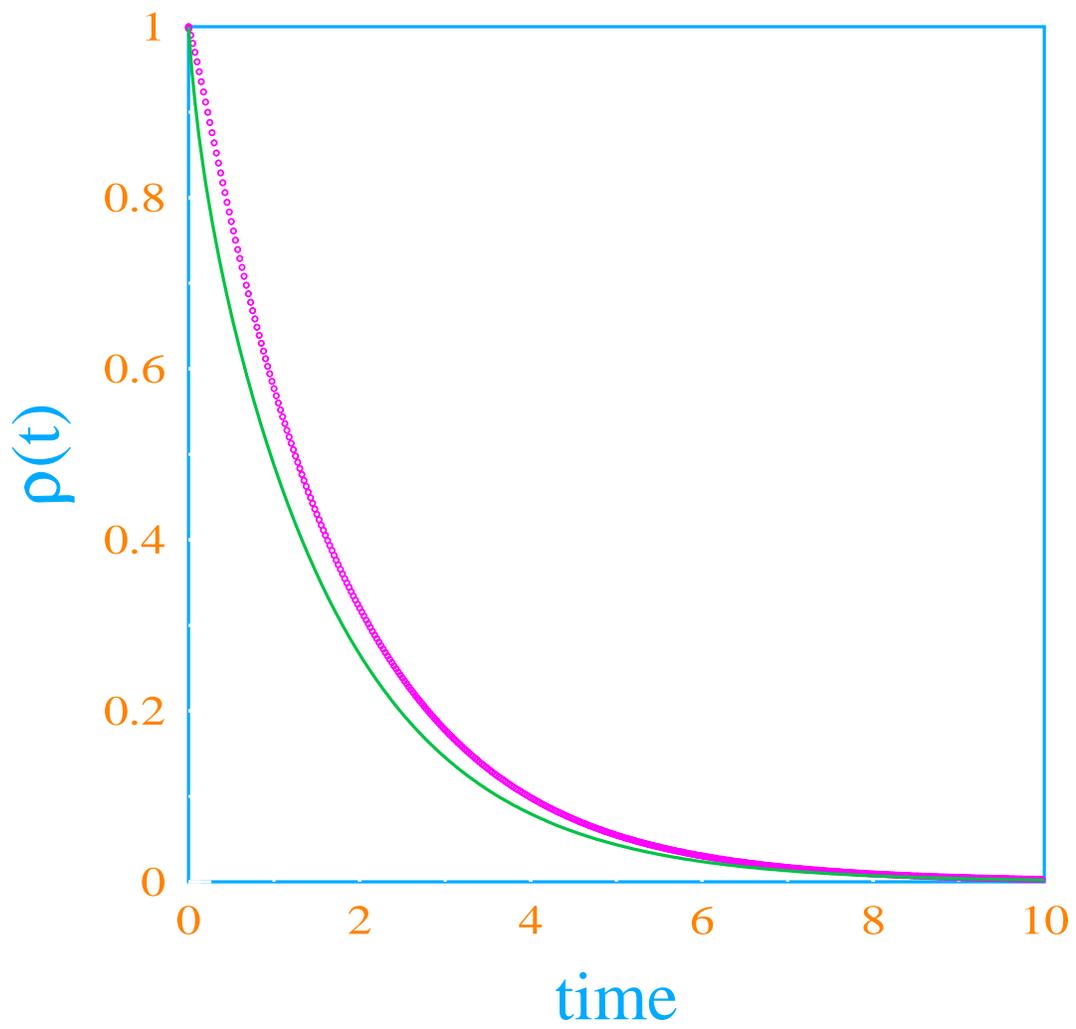
\end{document}